\pdfoutput=1

\documentclass[twocolumn,showkeys]{aastex631}
\usepackage{epstopdf}
\usepackage{xcolor}
\usepackage{xspace}
\usepackage{graphicx}
\usepackage{amssymb,amsmath}
\usepackage{natbib}
\usepackage{float}
\usepackage{hyperref}
\usepackage{mathptmx}
\usepackage{verbatim}
\usepackage{textcomp}
\usepackage{latexsym}
\usepackage{multirow}
\usepackage{wasysym}
\usepackage{academicons}
\usepackage{gensymb}
\usepackage{longtable}
\usepackage{ifthenx}
\shorttitle{Jet Termination Shock in Pictor\,A}
\shortauthors{Thimmappa et al.}

\accepted{2022 November 17}
\revised{2022 November 17} 
\received{2022 August 24}

\begin{document}
\nolinenumbers

\title{\textbf{X-ray Spectral Analysis of the Jet Termination Shock in Pictor\,A \\ on Sub-Arcsecond Scales with {\it Chandra}}}

\correspondingauthor{R.~Thimmappa}
\email{rameshan.thimmappa@villanova.edu}

\author[0000-0001-5122-8425]{R.~Thimmappa}
\affiliation{Villanova University, Department of Physics, Villanova, PA 19085, USA}

\author[0000-0001-8294-9479]{\L .~Stawarz}
\affiliation{Astronomical Observatory of the Jagiellonian University, ul. Orla 171, 30-244, Krak\'ow, Poland}

\author[0000-0002-8247-786X]{J.~Neilsen}
\affiliation{Villanova University, Department of Physics, Villanova, PA 19085, USA}

\author{\textcolor{xlinkcolor}{M.~Ostrowski} }
\affiliation{Astronomical Observatory of the Jagiellonian University, ul. Orla 171, 30-244, Krak\'ow, Poland}

\author[0000-0002-3778-1432]{B.~Reville} 
\affiliation{Max-Planck-Insitut f\"ur Kernphysik, Saupfercheckweg 1, Heidelberg 69117, Germany}

\begin{abstract}

Hotspots observed at the edges of extended radio lobes in high-power radio galaxies and quasars mark the position of mildly-relativistic termination shock, where the jet bulk kinetic energy is converted to the internal energy of the jet particles. These are the only astrophysical systems where mildly-relativistic shocks can be directly resolved at various wavelengths of the electromagnetic spectrum. The western hotspot in the radio galaxy Pictor\,A is an exceptionally good target in this respect, due to the combination of its angular size and high surface brightness. In our previous work, after a careful {\it Chandra} image deconvolution, we resolved this hotspot into a disk-like feature perpendicular to the jet axis, and identified this as the front of the jet termination shock. We argued for a synchrotron origin of the observed X-ray photons, which implied electron energies reaching at least 10–100\,TeV at the shock front.
Here we present a follow-up on that analysis, proposing in particular a novel method for constraining the shape of the X-ray continuum emission with sub-arcsec resolution. The method is based on a {\it Chandra} hardness map analysis, using separately de-convolved maps in the soft and hard X-ray bands. In this way, we have found there is a systematic, yet statistically significant gradient in the hardness ratio across the shock, such that the implied electron energy index ranges from $s\leq 2.2$ at the shock front to $s> 2.7$ in the near downstream. We discuss the implications of the obtained results for a general understanding of particle acceleration at mildly-relativistic shocks.

\end{abstract}

\keywords{radiation mechanisms: non--thermal --- galaxies: active --- galaxies: individual (Pictor A) -- galaxies: jets -- radio continuum: galaxies --- X-rays: galaxies}

\section{Introduction} 
\label{sec:intro}

Relativistic jets launched from high-accretion rate Active Galactic Nuclei (AGN), such as quasars and high-excitation radio galaxies, terminate by forming powerful shock waves, observed as prominent hotspots at the edges of extended radio cocoons/lobes inflated by the jets in the ambient medium \citep{Blandford74,Scheuer74}. In more detail, a light but high-power relativistic jet, when interacting with much denser interstellar/intergalactic medium, forms a double-shock structure: the non-relativistic forward shock propagates within the surrounding gas, compressing and heating the thermal plasma \citep[see, e.g.,][]{Carilli96,OSullivan18}, while the relativistic reverse shock converts the bulk kinetic energy of the outflow to the internal energy of jet particles \citep[e.g.,][]{Meisenheimer89,Kino04}. Magnetic field amplification and acceleration of some fraction of the jet particles to high, and even ultra-high energies, is expected to take place at the front of the reverse shock as well, although the exact acceleration processes, or the efficiency of the magnetic amplification, are still under the debate \citep[e.g.,][]{Stawarz07,Fan08,Araudo16,Araudo18,Matthews19}.

\begin{deluxetable*}{ccccccccc}[!th]
\tablecaption{Observational data and spectral fitting results for the soft ($0.5-2.0$\,keV) and hard ($2.5-7.0$\,keV) bands. \label{tab:PL_HR_map}}

\tablehead{\colhead{ObsID} & \colhead{Date} & \colhead{Exposure}  & Band & \colhead{Count rate} & \colhead{Photon index} & \colhead{$\chi^2/$dof} & \colhead{Energy flux} & \colhead{Net counts}\\
\colhead{} & \colhead{} & \colhead{[ksec]} & \colhead{} & \colhead{[cts/s]}  &  \colhead{$\Gamma$} & \colhead{} & \colhead{[$10^{-13}$\,erg\,cm$^{-2}$\,s$^{-1}$]} & \colhead{}}
	\startdata
	3090 & 2002-09-17 & 46.4 & soft & 0.078 & $1.90\pm 0.05$ & 67.91/100 & $2.69 \pm 0.05$ & 3,649 \\
                         & & & hard & 0.013 & $2.36\pm0.24$ & 27.89/52 & $2.00 \pm 0.07$ & 906 \\
	4369 & 2002-09-22 & 49.1 & soft  & 0.079 & $1.96\pm 0.05$ & 84.65/100 & $2.73 \pm 0.03$ & 3,894 \\
                         & & & hard & 0.018 & $2.35\pm 0.21$ & 30.94/55 & $1.96 \pm 0.04$ & 924\\
\enddata
\end{deluxetable*}

Hotspots in cosmologically distant radio quasars and high-power FR\,II radio galaxies are typically of the size of a few/several kiloparsecs, and so in order to study them properly, one needs instruments with at least arcsecond resolution. A considerable effort was made to resolve such structures at radio and infrared/optical frequencies, where hotspots shine through the synchrotron emission downstream of the reverse shock \citep[e.g.,][]{Prieto02,Brunetti03,Mack09,Perlman10,Orienti12,Orienti17,Orienti20,Pyrzas15,Dabbech18,Migliori20,Sunada22a}. Hotspots are also the sources of non-thermal X-ray photons, as established by numerous {\it Chandra} observations \citep{Hardcastle04,Kataoka05,Tavecchio05,Harris06,Massaro11,Massaro15,Mingo17}. The origin of the X-ray hotspots' emission is, in many cases, unclear: while in some sources the X-ray spectrum seems to fall into the extrapolation of the radio-to-optical synchrotron continuum, in other sources the X-ray excess suggests an additional emission component, typically ascribed to inverse-Comptonization of Cosmic Microwave Background photons, or of the hotspot's own synchrotron photons, by lower-energy electrons.

Among the other targets, the western (W) hotspot in the radio galaxy Pictor\,A is exceptionally well suited for deep observational studies, due to the combination of its relatively large angular size, very large angular separation from the bright galactic nucleus, and its high surface brightness. As such, it was subjected to a number of multiwavelength programs, including the radio domain with the Very Large Array \citep[VLA;][]{Perley97}, the mid-infrared range with the  IRAC camera onboard the {\it Spitzer} Space Telescope \citep{Werner12}, the Wide-field Infrared Survey Explorer \citep[WISE;][]{Isobe17}, and the SPIRE camera of the {\it Herschel} Space Observatory \citep{Isobe20}, at optical wavelengths with the Faint Object Camera on the {\it Hubble} Space Telescope \citep[HST;][]{Thomson95}, in X-rays with the Advanced CCD Imaging Spectrometer (ACIS) onboard the {\it Chandra} X-ray Observatory \citep{Wilson01,Hardcastle16,Thimmappa20}, as well as the EPIC MOS1 camera of the XMM-{\it Newton} \citep{Migliori07}, and lastly in hard X-rays with NuSTAR \citep{Sunada22b}. The hotspot was also the target of high-resolution radio imaging by the Very Long Baseline Array \citep[VLBA;][]{Tingay08}.

The radio structure of the W hotspot at GHz frequencies with the $1^{\prime\prime}.5$ VLA resolution is complex, including the main compact knot at the westernmost edge of the system, and the diffuse plateau region extending to the east/south-east \citep{Perley97}. With sub-arcsec VLA resolution (reaching $0^{\prime\prime}.17$), the main knot remains unresolved, while the plateau region reveals distinct filaments. The 74\,MHz---5\,GHz spectral index of the main knot is $\alpha \simeq 0.6 - 0.7$, and the degree of polarization reaches $70\%$; the upstream filaments seem to be characterized by a steeper spectrum ($\Delta \alpha \gtrsim 0.1$) and decreased polarization level (down to $10\%-30\%$). The projected magnetic field aligns with the levels of constant radio brightness, such that if the main compact knot denotes the position of the terminal reverse shock and its near downstream, the magnetic field configuration corresponds to that of a perpendicular shock. On the optical HST image with $\simeq 0^{\prime\prime}.1$ resolution \citep{Thomson95}, the main knot is decomposed into a system of highly polarized ($\gtrsim 50\%$) wisps elongated perpendicular to the jet axis.

Such complexity can hardly be followed at X-ray frequencies even with {\it Chandra}'s superb resolution. However, after a careful ACIS image deconvolution with sub-pixel resolution, presented in \citet{Thimmappa20}, the W hotspot could, in fact, be resolved into (i) a disk-like feature perpendicular to the jet axis, located $\simeq 1^{\prime\prime}.5$ to the south-east of the intensity peak of the main radio knot, but coinciding with the peak of the hotspot's optical emission, and (ii) an elongated feature aligned with the jet axis, and located even further upstream, i.e. within the region of the radio plateau. The disk-like feature could be traced for $\sim 4^{\prime\prime}$ in its longitudinal direction, but is resolved in its transverse direction only on sub-pixel scale.

\begin{figure*}[ht!]
	\centering 
	\includegraphics[width=0.49\textwidth]{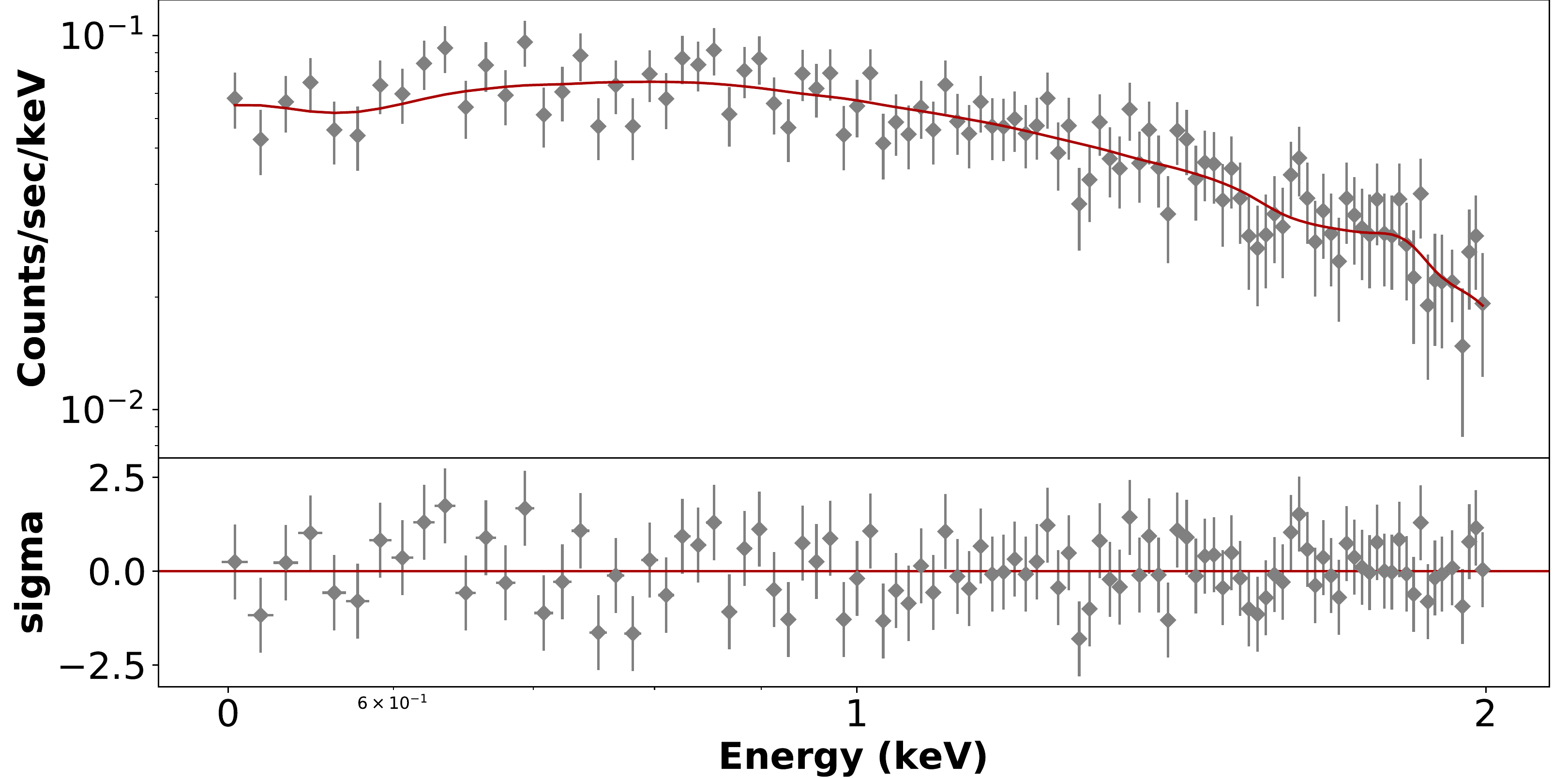}
	\includegraphics[width=0.49\textwidth]{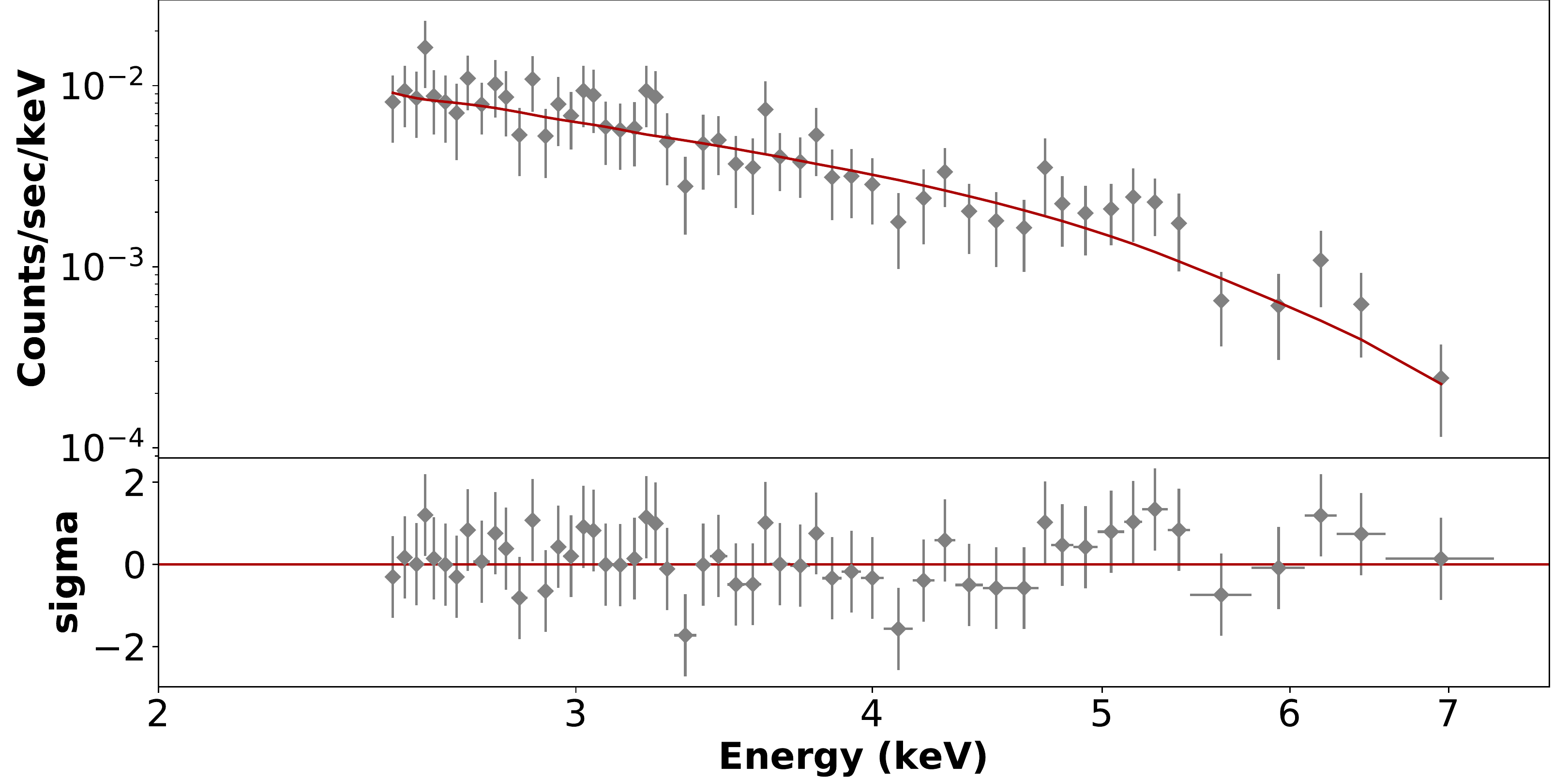}
	\includegraphics[width=0.49\textwidth]{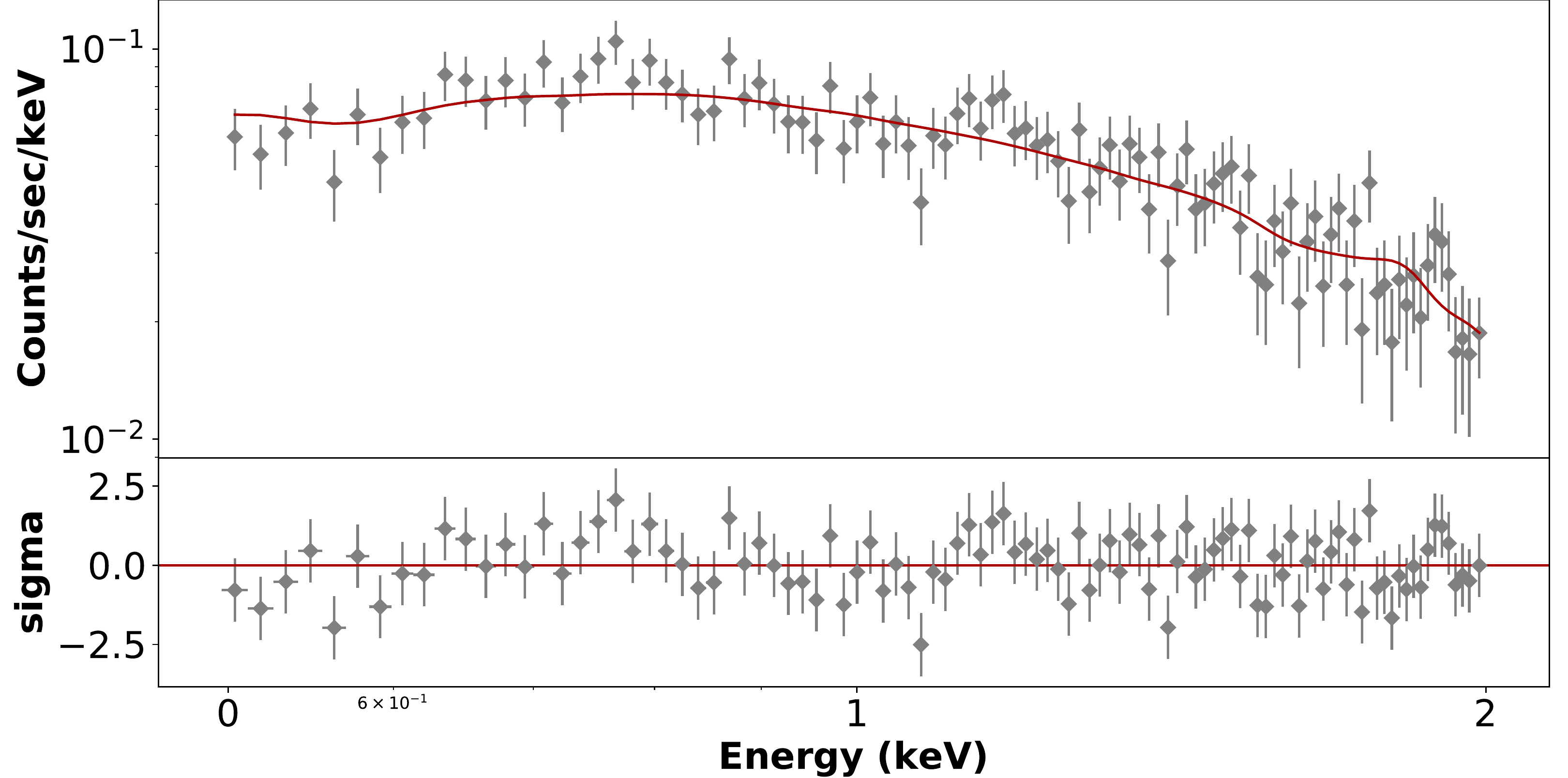}
	\includegraphics[width=0.49\textwidth]{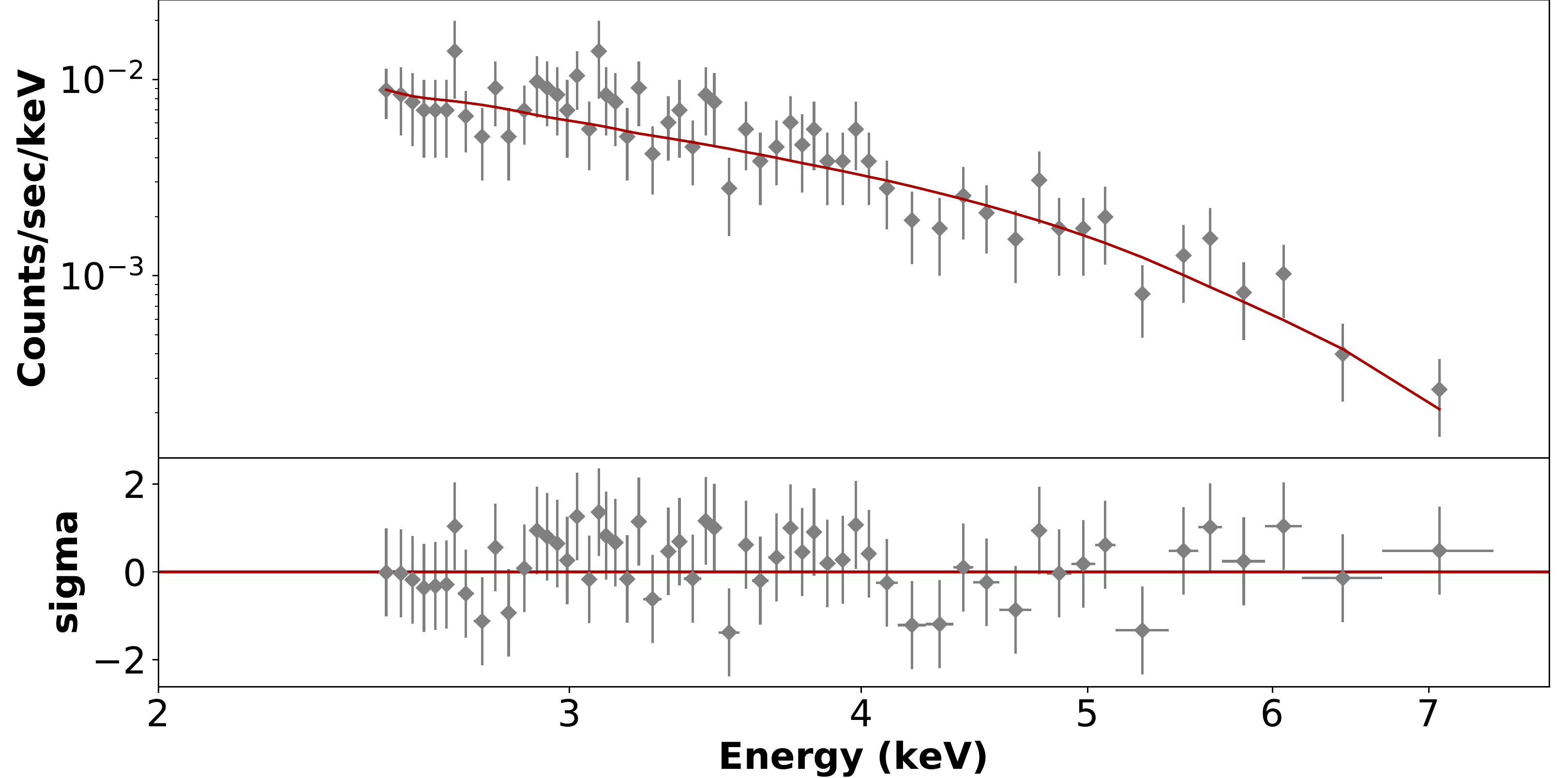}
	\caption{The $0.5-2.0$\,keV and $2.5-7.0$\,keV spectra of the entire (20\,px radius) W hotspot of Pictor\,A, for ID\,3090 (top) and 4369 (bottom), fitted with a single power-law model modified by the Galactic absorption (solid curves), along with the residuals.} 
\label{fig:PL_HR_map}
\end{figure*}

The overall interpretation of the observed multiwavelength morphology of the W hotspot therefore emerges, in which the perpendicular disk-like structure at the position of the hotspot's optical and X-ray intensity peaks corresponds to the very front of the reverse shock, where the most efficient particle acceleration is expected to take place. The radio intensity peak located further away, on the other hand, marks the downstream of the reverse shock, where the radiative cooling of the plasma convected away from the shock front prevents production of high-energy optical and X-ray synchrotron photons. Finally, the nature of the X-ray jet-like feature upstream of the shock, as well as optical and radio filaments within the extended plateau region, remain unclear, although such structures may be related to a network of weaker oblique shocks formed around the head of the jet by the plasma back-flowing from the downstream of the reverse shock \citep[see, e.g.,][]{Saxton02,Mizuta10}.

The good agreement between the optical and X-ray maps, along with the general X-ray spectral properties of the hotspot, as well as hints for the X-ray time variability of the target, all imply in accord synchrotron origin of the observed X-ray photons \citep{Hardcastle16,Thimmappa20,Sunada22b}. For this reason, the X-ray spectral properties of the hotspot are crucial for a proper understanding of particle acceleration processes taking place at mildly-relativistic perpendicular shocks in general. And indeed, the very presence of X-ray synchrotron photons means that such shocks are able to accelerate electrons up to energies $E_e \sim 10^8 \, m_ec^2$, assuming the hotspot magnetic field of the order of $B \sim 0.1-1$\,mG \citep[see the discussion in][]{Thimmappa20,Sunada22b}. 

However, for an X-ray spectral analysis with any of the available X-ray instruments, the source extraction region has to be relatively large, in order to maximize the photon statistics for a given Point Spread Function (PSF) and the source intrinsic extent. Such an extended region unavoidably includes therefore various sub-components of the system, and hence the resulting spectral constraints do not correspond to the reverse shock exclusively, but instead to a superposition of the reverse shock, its downstream region, and also of the upstream filaments/jet-like features. Here we propose a novel, alternative method for constraining the shape of the X-ray continuum emission at the very position of the reverse shock, with sub-arcsec resolution. The method is based on hardness map analysis, for \emph{separately de-convolved} soft and hard maps; this novelty resolves the problem of artefact features appearing on X-ray hardness maps due to the energy-dependent {\it Chandra} PSF. 

Throughout the paper we assume $\Lambda$CDM cosmology with $H_{0} = 70$\,km\,s$^{-1}$, $\Omega_{\rm m} = 0.3$, and $\Omega_{\Lambda} = 0.7$. The Pictor\,A redshift $z=0.035$ \citep{Eracleous04}, corresponds therefore to the luminosity distance of 154\,Mpc, and the conversion angular scale 0.7\,kpc\,arcsec$^{-1}$. The photon index $\Gamma$ is defined here as $F_{\varepsilon} \propto \varepsilon^{-\Gamma}$ for the photon flux spectral density $F_{\varepsilon}$ and the photon energy $\varepsilon$; the spectral index is $\alpha = \Gamma - 1$.

\begin{figure*}[ht!]
	\centering 
	\includegraphics[width=\textwidth]{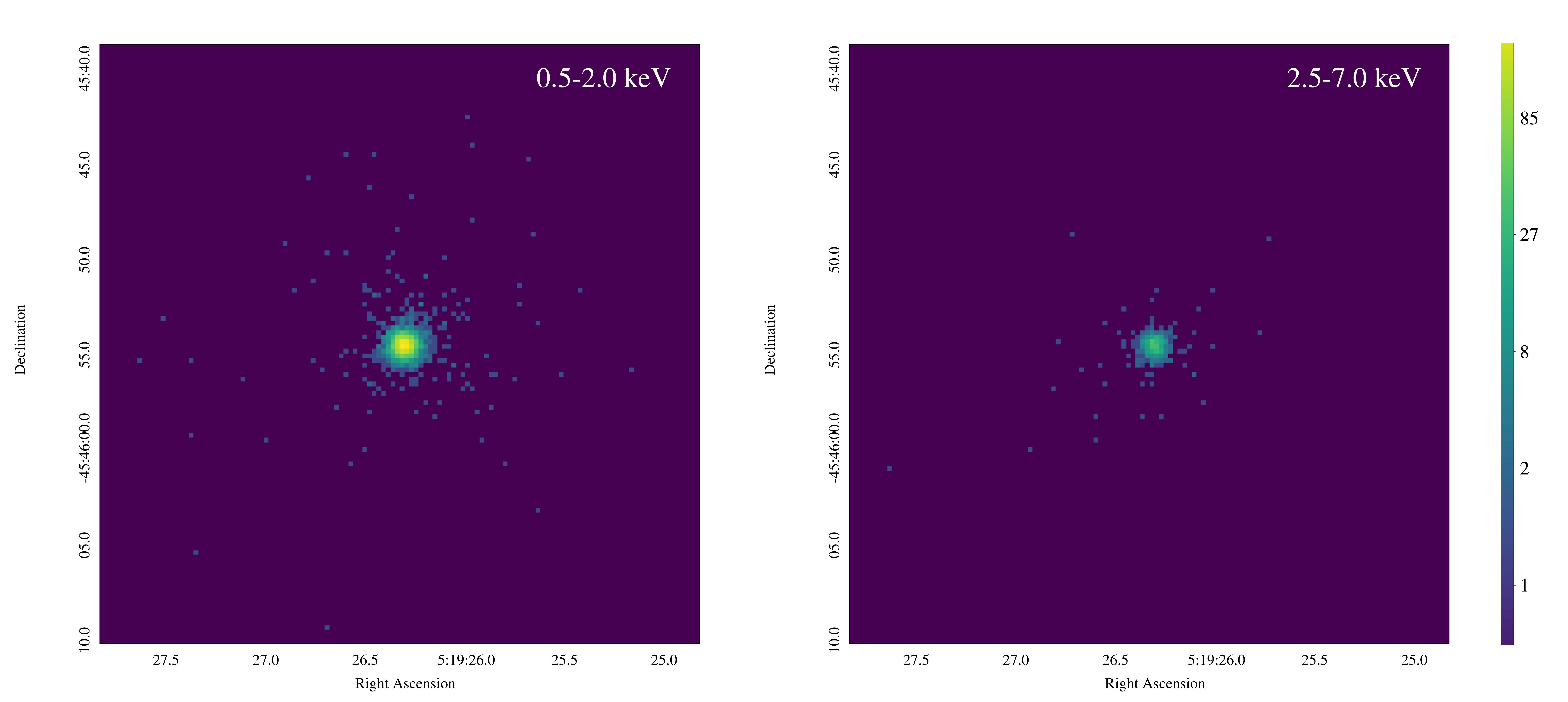}
	\caption{Examples of simulated PSF images for the ObsID\,3090 in the soft and hard bands (left and right panels, respectively), with 0.5\,px resolution.}
	\label{fig:psf} 
\end{figure*}

\section{Chandra Data} 
\label{sec:data}

The W hotspot of Pictor\,A was observed on-axis with the ACIS \citep{Garmire03} onboard the {\it Chandra} X-ray Observatory \citep{Weisskopf00} on 2002-09-17 (ID\,3090) and 2002-09-22 (ID\,4369). A combination of relatively long uninterrupted exposures for both pointings, totaling to an observing time of 95.5\,ksec, and a small off-axis angles $\theta \simeq 0.^{\prime}11$, makes them ideal dataset for our high-resolution study. 

The observational data were reprocessed using the {\ttfamily chandra\_repro} script as per the Chandra Interactive Analysis of Observations \citep[CIAO v4.14;][]{Fruscione06} analysis threads\footnote{\url{http://cxc.harvard.edu/ciao/threads/}}, using Chandra Calibration Database (CALDB)\,v4.9.7. Pixel randomization and readout streaks were removed from the data during processing. Point sources in the vicinity of the hotspot were detected with {\ttfamily wavdetect} tool using the minimum PSF method, and removed. For our analysis, we selected photons in the range $0.5-7.0$\,keV. Photon counts and spectra were extracted for the source and background regions from individual event files using the {\ttfamily specextract} script. Spectral fitting was done with the {\fontfamily{qcr}\selectfont Sherpa} package \citep{Freeman01}.

The total number of counts for the hotspot, $\sim 10,000$ for both exposures together (see Table\,\ref{tab:PL_HR_map}), places us in the regime where calibration uncertainties dominate over statistical uncertainties \citep{Drake06}. Methods to account for calibration uncertainties in the analysis of {\it Chandra} data have been discussed by \citet{Lee11} and \citet{Xu14}. The moderate count-rate of $\simeq 0.1$\,s$^{-1}$ for the hotspot located at the center of the S3 chip, implies only small chances for a pile-up in the detector \citep{Davis01}; we have verified it during the spectral analysis, but nonetheless have included the pile-up model when performing {\fontfamily{qcr}\selectfont MARX} simulations anyway (see the following section).

\begin{figure*}[ht!]
	\centering 
	\includegraphics[width=0.49\textwidth]{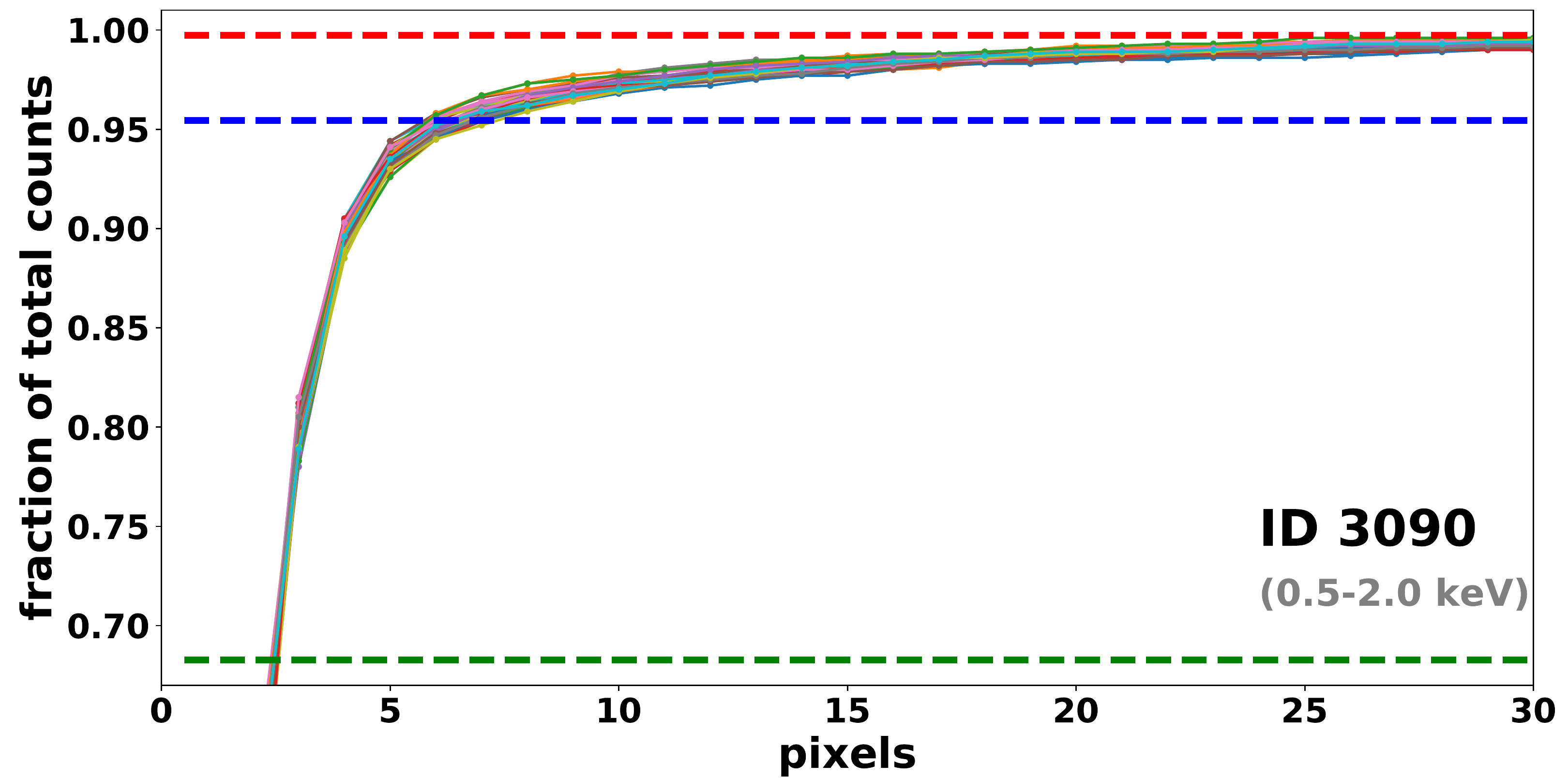}
	\includegraphics[width=0.49\textwidth]{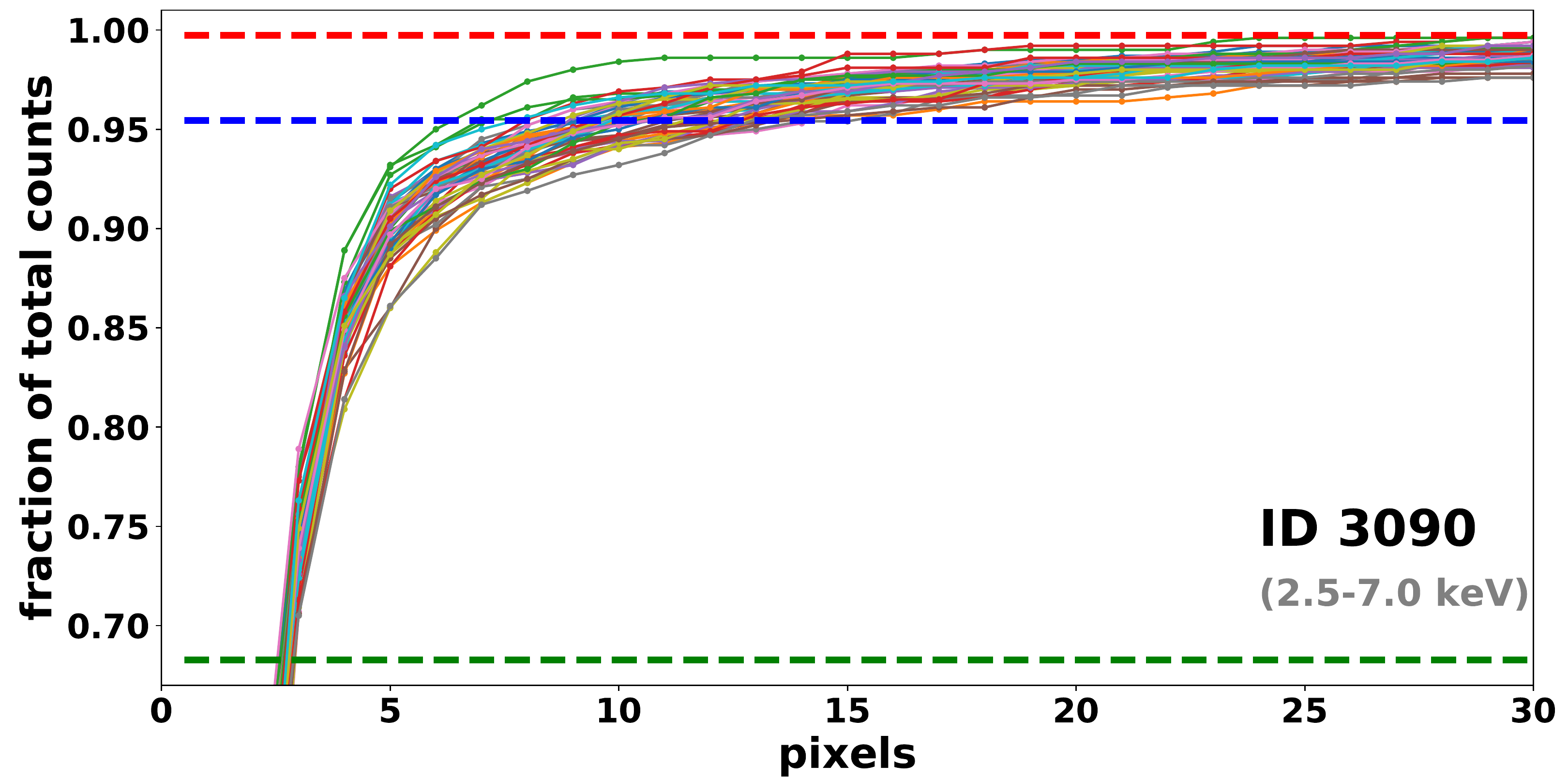}
	\caption{Enclosed count fraction (ECF) as a function of the aperture radius for all 50 the simulated PSFs in the soft and hard bands (left and right panels), ObsID\,3090. The horizontal green, blue, and red lines correspond to $1 \sigma$, $2\sigma$, and $3 \sigma$ ECFs, respectively.}
	\label{fig:ecf} 
\end{figure*}

\section{Data Analysis} 
\label{sec:analysis}

\subsection{Spectral Modeling}
\label{sec:spectrum}

A composite hotspot spectrum was extracted for each ObsID from a circular region (position: RA\,=\,5:19:26.2993, DEC\,=\,--45:45:54.377) with a radius 20\,px ($\simeq 10^{\prime\prime}$, for the conversion scale $0.492^{\prime\prime}/{\rm px}$), and the background set as a concentric annulus of 30-60\,px radius \citep[see][Figure\,1 therein]{Thimmappa20}. The background-subtracted hotspot spectra were next fitted within the soft ($0.5-2.0$\,keV) and hard ($2.5-7.0$\,keV) bands \emph{separately}, assuming a power-law model modified by the Galactic column density $N_{\rm H,\,Gal} = 3.6 \times 10^{20}$\,cm$^{-2}$ \citep{HI4PI2016}. The results of spectral fitting are summarized in Table\,\ref{tab:PL_HR_map}, and the fitted spectra are shown in Figure\,\ref{fig:PL_HR_map}. 

The power-law models with photon indices $\Gamma \simeq 1.9$ in the soft band, and significantly larger $\Gamma \sim 2.7$ in the hard band, provide a reasonable description of the source composite spectra, sufficient in particular for the purpose of the PSF modeling. We note that analogous fits with the Galactic absorption set free returned similar results, only with slightly decreased values of $N_{\rm H,\,Gal}$ and $\Gamma$. Finally, including the {\ttfamily jdpileup} model in the fitting procedure does not affect the best-fit values of the model parameters, as the fraction of piled-up events that result in a good grade turns out to be very low.

\subsection{PSF Modeling}
\label{sec:PSF}

To model the {\it Chandra} PSF at the position of the W hotspot, we used the Chandra Ray Tracer ({\fontfamily{qcr}\selectfont ChaRT}) online tool \citep{Carter03}\footnote{\url{http://cxc.harvard.edu/ciao/PSFs/chart2/runchart.html}} and the {\fontfamily{qcr}\selectfont MARX} software \citep{Davis12} \footnote{\url{https://space.mit.edu/cxc/marx}}. For both ObsID\,3090 and 4369, the centroid coordinates of the selected source region were taken as the position of a point source. The source spectra for {\fontfamily{qcr}\selectfont ChaRT} were the respective power-law models in the 0.5--2.0\,keV and 2.5--7.0\,keV bands, as described in Section\,\ref{sec:spectrum}. Since each particular realization of the PSF is different due to random photon fluctuations, in each case a collection of 50 event files was made, with 50 iterations using {\fontfamily{qcr}\selectfont ChaRT} by tracing rays through the {\it Chandra} X-ray optics. The rays were projected onto the detector through {\fontfamily{qcr}\selectfont MARX} simulation, taking into account all the relevant detector effects, including pileup and energy-dependent sub-pixel event repositioning. The PSF images were created with the size of $32\times 32$\,pix$^2$, and binned with 0.5\,px resolution. An example of the simulated PSF images for ObsID\,3090 in the soft and hard bands is presented in Figure\,\ref{fig:psf}.

In order to illustrate the size of the PSFs in both bands, we calculated the enclosed count fraction (ECF) for all the simulated PSFs, i.e., the fraction of counts that would be detected within a certain circular aperture for a particular realization of the PSF. The resulting ECFs are presented in Figure\,\ref{fig:ecf} for the soft and hard bands (left and right panels, respectively), ObsIDs\,3090. As shown, for the soft band, the $2 \sigma$ radius is typically $\simeq 6$\,px, while in the hard band the corresponding $2 \sigma$ radius has a wider spread, ranging from $\simeq 6$\,px for some realizations of the PSF, up to even $\simeq 15$\,px.

\begin{figure*}[ht!]
	\centering 
	\includegraphics[width=\textwidth]{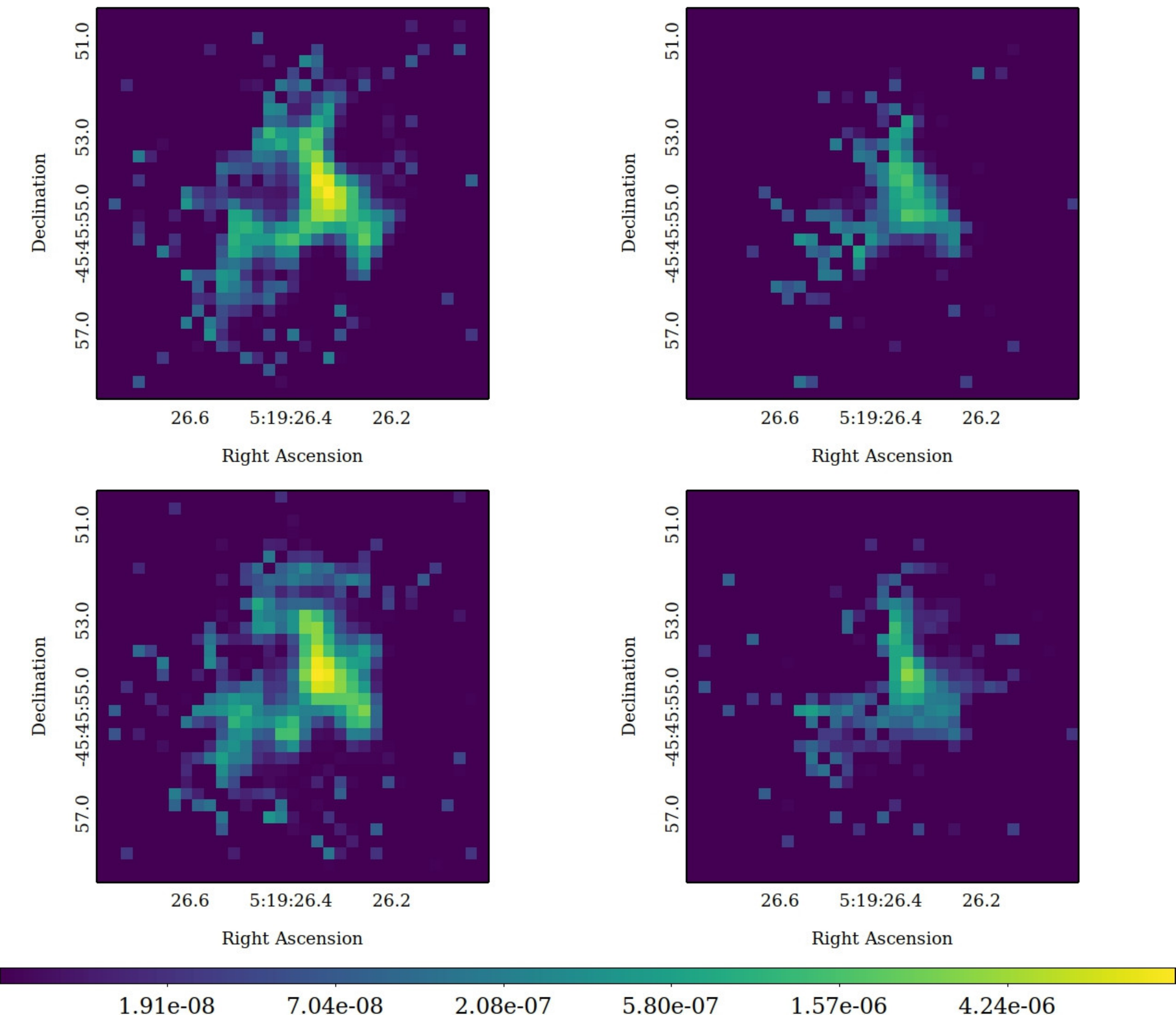}
	\caption{Deconvolved images of the W hotspot in Pictor\,A, in the soft 0.5--2.5\,keV and hard 2.5--7.0\,keV bands (left and right panels, repectively), for the ObsID\,3090 and ObsID\,4369 (top and bottom panels, respectively). Each image corresponds to the average over 50 PSF simulations with 0.5\,px resolution.}
	\label{fig:deconv} 
\end{figure*}

\begin{figure*}[ht!]
	\centering 
	\includegraphics[width=\textwidth]{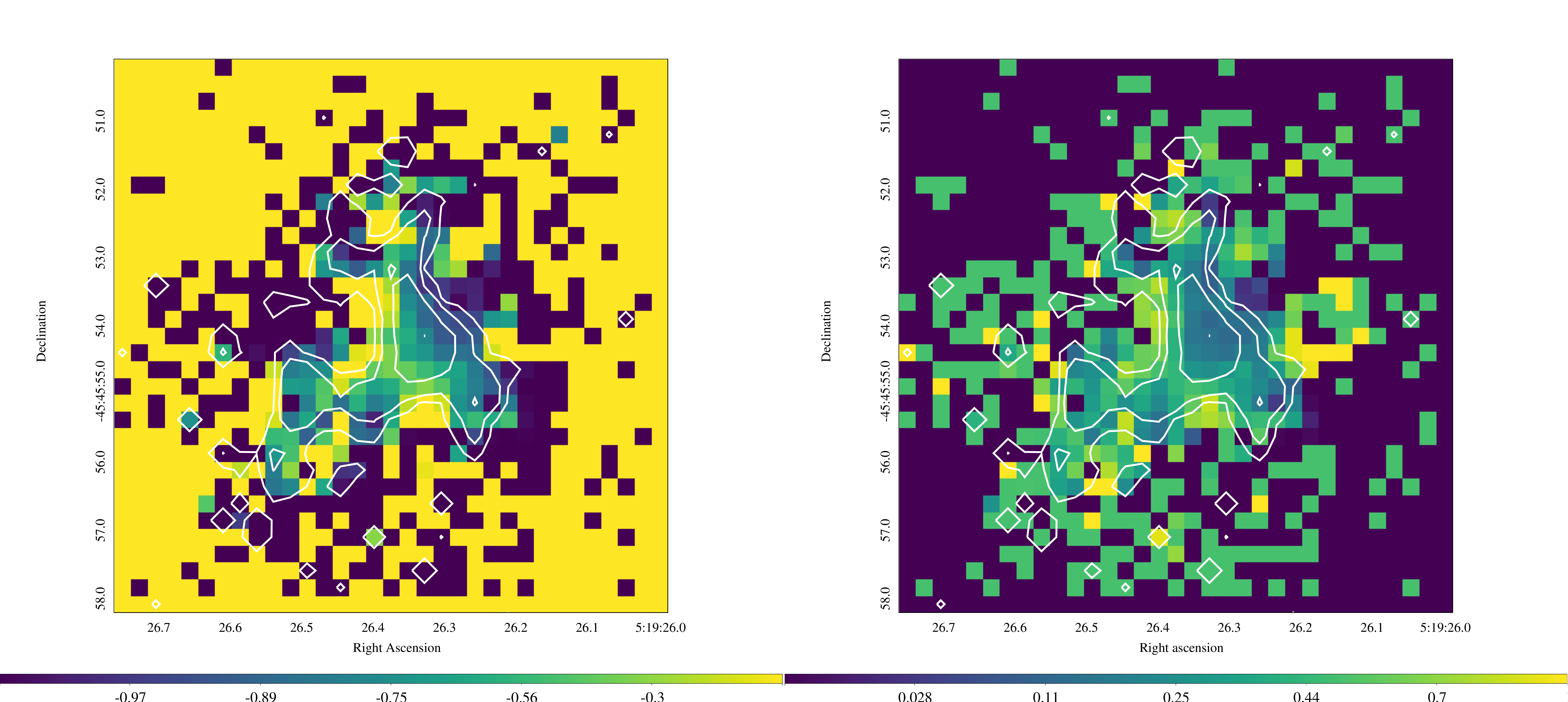}
	\caption{{\it Left panel:} Hardness ratio map of the W hotspot in Pictor\,A, corresponding to the combined de-convolved {\it Chandra} images with 0.5\,px resolution (combined observations ID\,3090 and 4369); white contours denote the total {\it Chandra} intensity, based similarly on de-convolved images with 0.5\,px resolution. {\it Right panel:} The corresponding standard deviation ($\sigma$) of the hardness ratio value, again with the total intensity contours superimposed.}
	\label{fig:HR} 
\end{figure*}

Note that, since the region encompassing the hotspot structure is relatively compact, one should not expect any significant change of the PSF across the field subjected to the image deconvolution, as described in the next section. The spectral information provided for the PSF modeling, on the other hand, corresponds to the composite radiative output of the entire structure, while below we argue for the presence of significant spectral changes on sub-pixel scale within the brightest segments of the hotspot. This inherent inconsistency does not however affect the main results of the analysis.

\subsection{Image Deconvolution}
\label{sec:deconvolve}

We used the Lucy-Richardson Deconvolution Algorithm (LRDA), which is implemented in the {\fontfamily{qcr}\selectfont CIAO} tool {\ttfamily arestore}, to remove the PSF blurring, and in this way to restore the intrinsic surface brightness distribution of the hotspot. This method does not affect the number of counts on the image, but only their distribution.

The algorithm requires an image form of the PSF, which is provided by our {\fontfamily{qcr}\selectfont ChaRT} and {\fontfamily{qcr}\selectfont MARX} simulations as described in Section\,\ref{sec:PSF} above, and exposure-corrected maps of the source \citep[for more details see][]{Marchenko17,Thimmappa20}. Here we perform the de-convolution separately for the soft and hard bands, in each case for 50 random realizations of the simulated PSF; those 50 deconvolved images were then averaged to a single image using {\ttfamily dmimgcalc} tool. The resulting images are shown in Figure\,\ref{fig:deconv}. The two main features of the hotspot observed by \citet{Thimmappa20}, namely the disk-like feature perpendicular to the jet axis, as well as a weaker jet-like feature extending to the south-east along the jet axis, are present on both soft and hard maps for both ObsIDs, although the jet-like feature is much less prominent on the hard maps.

\subsection{Hardness Ratios}
\label{sec:HR maps}

Based on the de-convolved soft and hard images of the W hotspot in Pictor\,A with 0.5\,px resolution, we perform spatially-resolved Hardness Ratio (HR) analysis, in order to investigate the spectral structure of the system on sub-arcsec scales, free, as much as possible, from the PSF blurring. 

\begin{figure*}[th!]
	\centering 
	\includegraphics[width=0.53\textwidth]{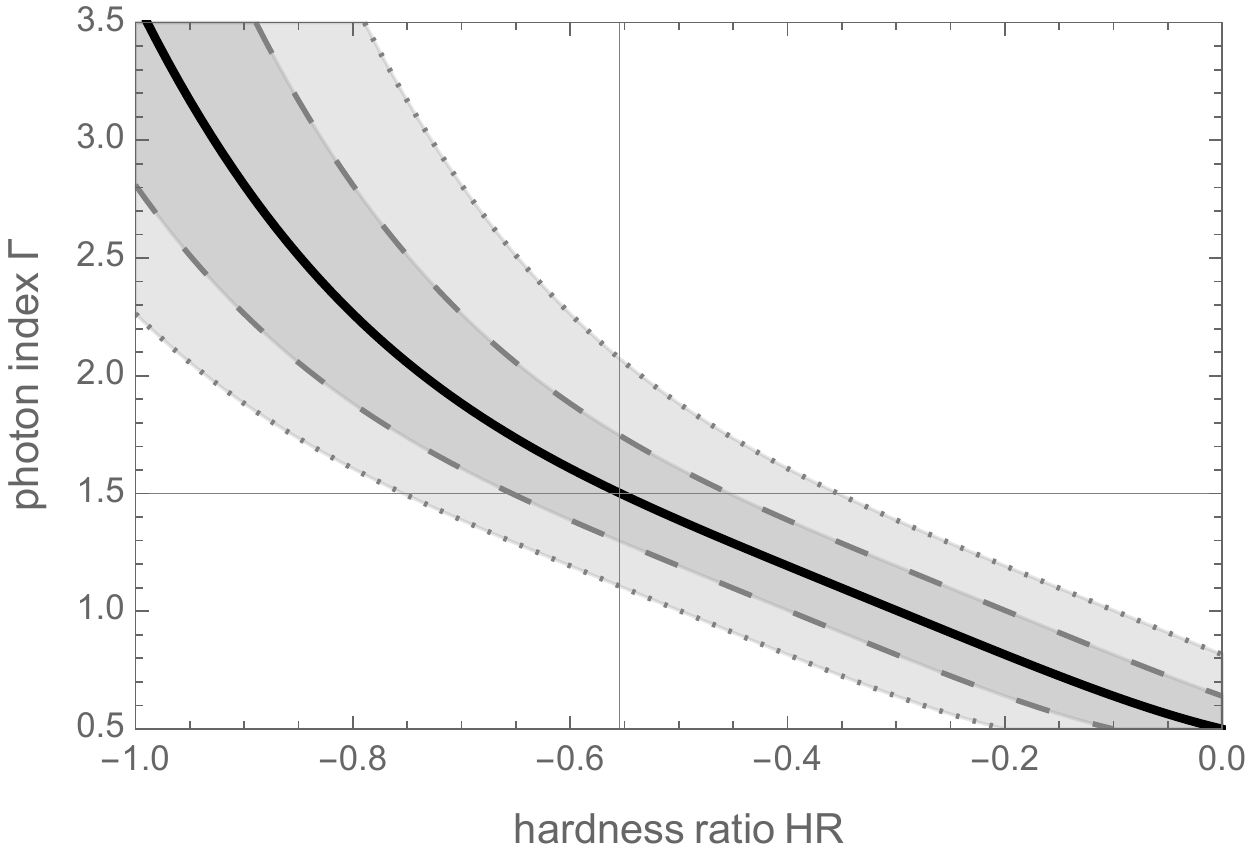}
    \includegraphics[width=0.435\textwidth]{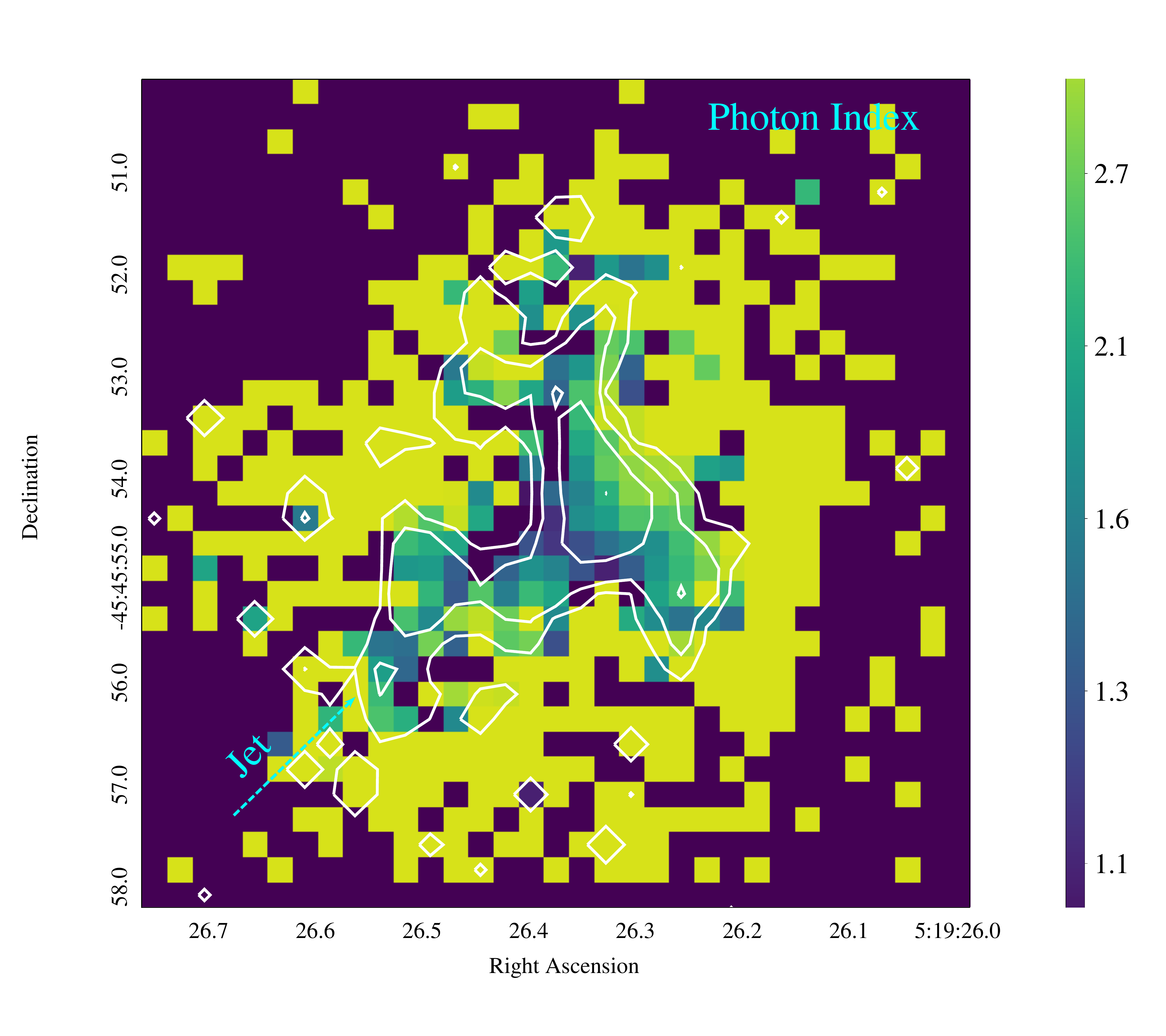}
   	\caption{{\it Left panel:} Black solid curves denote the values of the photon index $\Gamma$, corresponding to different values of the 0.5--7.0\,keV hardness ratio HR, assuming Galactic column density $N_{\rm H,\,Gal} = 3.6 \times 10^{20}$\,cm$^{-2}$. Dark-gray and gray regions in between the dashed and dotted curves, denote respectively the $\pm 0.1$ and $\pm0.2$ statistical uncertainty in HR. The ``canonical'' shock-type index $\Gamma=1.5$, corresponds to the HR value of $\simeq -0.56$ (thin solid lines in the panel). {\it Right panel:} The resulting 0.5--7.0\,keV photon index map corresponding to the 0.5\,px resolution HR map presented in the left panel of Figure\,\ref{fig:HR}, obtained assuming Galactic column density $N_{\rm H,\,Gal} = 3.6 \times 10^{20}$\,cm$^{-2}$.}
	\label{fig:indices} 
\end{figure*}

HR analysis of {\it Chandra} data has been widely applied to various classes of astrophysical sources before \citep[e.g.,][]{Balucinska05,Siemiginowska07,Nandra15,Haggard19}, being in particular considered as a useful tool that allows constraints on spectra of unresolved weak sources, for which the standard spectral modeling approach is not possible due to low numbers of counts. Spatially-resolved HR analysis for extended sources, however, remains largely unexplored, because of artefact features appearing on the HR maps, in relation to (i) the energy dependence of the {\it Chandra}'s PSF, and also (ii) random fluctuations of photons, relevant especially in the low surface brightness regime. Our approach resolves the aforementioned problems, since (i) the HR analysis is based on the separately de-convolved soft and hard maps, and (ii) we remove the effect of random fluctuations by averaging over 50 realizations of each modelled PSF. In particular, based on the soft (S) and hard (H) de-convolved images, we produce spectral maps defined as ${\rm HR}=({\rm H}-{\rm S})/({\rm H}+{\rm S})$. Next, we average the HR maps corresponding to ObsID\,3090 and 4369, obtaining at the end the final distribution of the X-ray HR for the W hotspot of Pictor\,A, shown in the left panel of Figure\,\ref{fig:HR}.

An HR variance map, i.e. the values of the variance at a given position $(x,y)$ on the map, was generated based on the same $N=100$ deconvolved HR images (50 for ObsID\,3090 and another 50 for ObsID\,4369), ${\rm HR}_i(x,y)$, simply as
\begin{equation}
V(x,y) = \frac{1}{N-1} \, \sum_{i=1}^{N} \left[{\rm HR}_i(x,y)-\overline{{\rm HR}(x,y)}\right]^2 \, ,
\end{equation}
where $\overline{{\rm HR}(x,y)}$ denotes the averaged HR image. The square-root of this variance, $\sigma = \sqrt{V(x,y)}$, corresponds therefore to the \emph{statistical} uncertainty in the derived values of the hardness ratio at a given position $(x,y)$ on the map. This uncertainty is shown in the right panel of Figure\,\ref{fig:HR}.

The main structure of the hotspot that is prominent on the total intensity map is characterized by the values $-1<$\,HR\,$<0$. This structure is surrounded by a soft halo with ${\rm HR}=-1$, meaning simply no hard photons; outside of the soft halo, where the X-ray flux also drops in the soft band, the HR values fluctuate around 0. This assures the reality of the spectral map produced, as no artifact features are present in the regions with background-level flux, and all the physically meaningful HR values are concentrated exclusively in the high-flux region. Moreover --- and this is the major finding of the analysis --- there is a clear systematic HR gradient across the main disk-like feature, ranging from approximately $-0.4$ down to $-0.9$ and below (see the left panel of Figure\,\ref{fig:HR}) across the main disk-like feature, i.e. from the upstream (south-east) to the downstream (north-west) of the shock. The HR uncertainty in that region is on average, $\pm 0.2$ (see the right panel in Figure\,\ref{fig:HR}), so that the HR gradient is statistically significant.

In addition to the statistics, however, a careful investigation of the systematic uncertainty is also required. We have therefore produced hardness maps of other astrophysical sources appearing point-like for {\it Chandra}, in the analogous way as described above. For a fair comparison with the Pictor\,A hotspot, we selected sources that are unresolved and were observed by {\it Chandra} around 2002 (in order to avoid complications related to the CCD degradation), were not variable during the {\it Chandra} exposure, were free of pile-up, and had comparable photon statistics to those of the analyzed Pictor\,A pointings. The best targets fulfilling such criteria, were the BL\,Lac object AO\,0235+16 ($z = 0.94$), and quasar 4C+13.85 ($z=0.673$). In both cases, we 
found no evidence of any substructure introduced by the hardness ratio map. Thus we do not believe that the gradient seen above is a systematic effect of our method. The corresponding maps for the two targets are presented in Appending~\ref{A:comparative}.

\section{Discussion and Conclusions} 
\label{sec:results} 

Assuming a single power-law emission model, a given value of the HR corresponds to a particular set of values for the photon index $\Gamma$ and Galactic column density $N_{\rm H,\,Gal}$ (assuming zero intrinsic absorption). In Figure\,\ref{fig:indices}, left panel, we plot this dependence, adopting $N_{\rm H,\,Gal} = 3.6 \times 10^{20}$\,cm$^{-2}$. With such, the value HR\,$=-0.4$ gives the photon index $\Gamma \simeq 1.2$, while for example HR\,$=-0.9$ gives $\Gamma \simeq 2.8$. The resulting 0.5--7.0\,keV photon index map of the W hotspot in Pictor\,A, corresponding to the 0.5\,px resolution HR map discussed above, is given in the right panel of Figure\,\ref{fig:indices}. The uncertainties in the exact $N_{\rm H,\,Gal}$ value, even if at the level of $\sim 50\%$, are in this context much less relevant than the statistical HR mean uncertainty of $\simeq 0.2$, following from the square-root variance mapping of the disk feature. This statistical uncertainty would in particular translate into a wider range of the allowed photon indices, roughly speaking $\Gamma \leq 1.6$ for the upstream edge, and $\Gamma \geq 1.9$ for the downstream region. In the case of the synchrotron origin of the detected X-ray photons, those values of photon indices would then correspond to the index of the electron energy distribution $s \equiv - \log N\!(E_e) / \log E_e = 2 \, \Gamma - 1$ ranging from $\leq 2.2$ up to $>2.8$. 

It is however important to emphasize at this point, that the \emph{broad-band} spectrum of ultra-relativistic electrons accelerated at the shock front, may be much more complex than a single power-law. A single power-law model is used here rather for illustrative purposes, to give a basic insight into the slope of the high-energy segment of the electron distribution, and the amount of spectral steepening observed across the shock front in the Pictor\,A hotspot.

The gradient in the HR values across the terminal reverse shock we have found has several important implications for understanding particle acceleration at relativistic shocks in general. First, the fact that the hardest X-ray spectra we see are concentrated at the upstream edge of the X-ray intensity peak, means that the efficient electron acceleration --- forming flat electron energy distributions with indices $s\leq 2.2$ (when approximated by a single power-law) and electron energies of the order 10-100\,TeV --- takes place at the very front of the \emph{mildly-relativistic shock with perpendicular magnetic field configuration}, and not in the far downstream, for example, where compact radio knots have been found in VLBA observations \citep{Tingay08}. Second, the HR gradient suggests that high-energy electrons advected from the shock front cool radiatively (leading to a steepening of their energy distribution and the corresponding X-ray spectrum). This confirms the origin of the offset between the X-ray hotspot and the VLA hotspot: the propagation length of the ultrarelativistic electrons that produce keV photons is of the order of a parsec for the expected hotspot magnetic field 0.1--1\,mG, and at most a hundred of parsecs for unrealistically low magnetic field intensity of a few $\mu$G \citep{Thimmappa20}, while it is much longer for radio-emitting electrons. By the time the jet has traveled the $\simeq 1$\,kpc between the X-ray hotspot and the compact radio knots, there are essentially no X-ray emitting electrons left.

Mildly-relativistic magentized shocks in electron--ion plasmas --- meaning shock bulk Lorentz factors $\gamma_{\rm sh} \lesssim (m_p/m_e)^{1/3} \simeq 10$ and magnetization parameters, defined as a ratio of the upstream Poynting flux to the kinetic energy flux, $10^{-3} < \sigma \leq 0.1$, matching the conditions expected to hold in the western hotspot of Pictor\,A --- have been investigated with 2D kinetic particle-in-cell simulations by \citet{Sironi11}, and more recently by \citet{Ligorini21a,Ligorini21b}. These studies do show some energization of electrons, due to the resonant interactions with large-amplitude longitudinal Langmuir waves, combined with shock-surfing acceleration \citep{Lyubarsky06,Hoshino08}, however with a rather low efficiency when compared to ultra-relativistic shocks (i.e., shocks with $\gamma_{\rm sh} \gg 10$). As a consequence, the downstream electron spectra observed in such simulations (i) are basically thermal with little or no non-thermal power-law components, (ii) have total energy density much below that of the ions, at the level of about $10\%$, and (iii) have limited maximum energies $E_e/m_e c^2 < (m_p/m_e) \gamma_{\rm sh} < 10^4$. This is in contrast to the observational findings presented here, and elsewhere in the literature, regarding the ion-electron energy equipartition \citep[see][]{Stawarz07}, electron energies of the order of 10-100\,TeV \citep[see][]{Sunada22b}, and flat slopes of the electron energy distribution (this work). Together, these observational findings indicate that electron acceleration is both fast and efficient at the jet termination shocks of luminous radio galaxies and quasars.

\begin{acknowledgements}

This research has made use of data obtained from the Chandra Data Archive. This work was supported by the Polish NSC grant 2016/22/E/ST9/00061 (R.T., \L .S.) and NASA award 80NSSC20K0645 (R.T., J.N). The authors thank the anonymous referee and O.~Kobzar for valuable comments and suggestions on the manuscript.

\end{acknowledgements}

\appendix
\section{Hardness Ratio Analysis of the Comparative Sources}
\label{A:comparative}

BL Lac object AO\,0235+16 ($z = 0.94$), has been observed on 20-08-2000 by {\it Chandra} on the ACIS-S3 chip (ObsID\,884) with 30.625\,ksec exposure time. The source spectrum was extracted from a circular region (position: RA\,=\,2:38:38.9560, DEC\,=\,16:36:59.440) with a radius 3\,px ($\simeq 1.5^{\prime\prime}$), and the 5-10\,px annulus background. The background-subtracted source spectra were fitted within the soft ($0.5-2.0$\,keV) and hard ($2.5-7.0$\,keV) bands, assuming single power-law models modified by the Galactic column density $N_{\rm H,\,Gal} = 6.79\times 10^{20}$\,cm$^{-2}$ \citep{HI4PI2016}.

\begin{figure}[ht!]
	\centering
	\includegraphics[width=0.89\textwidth]{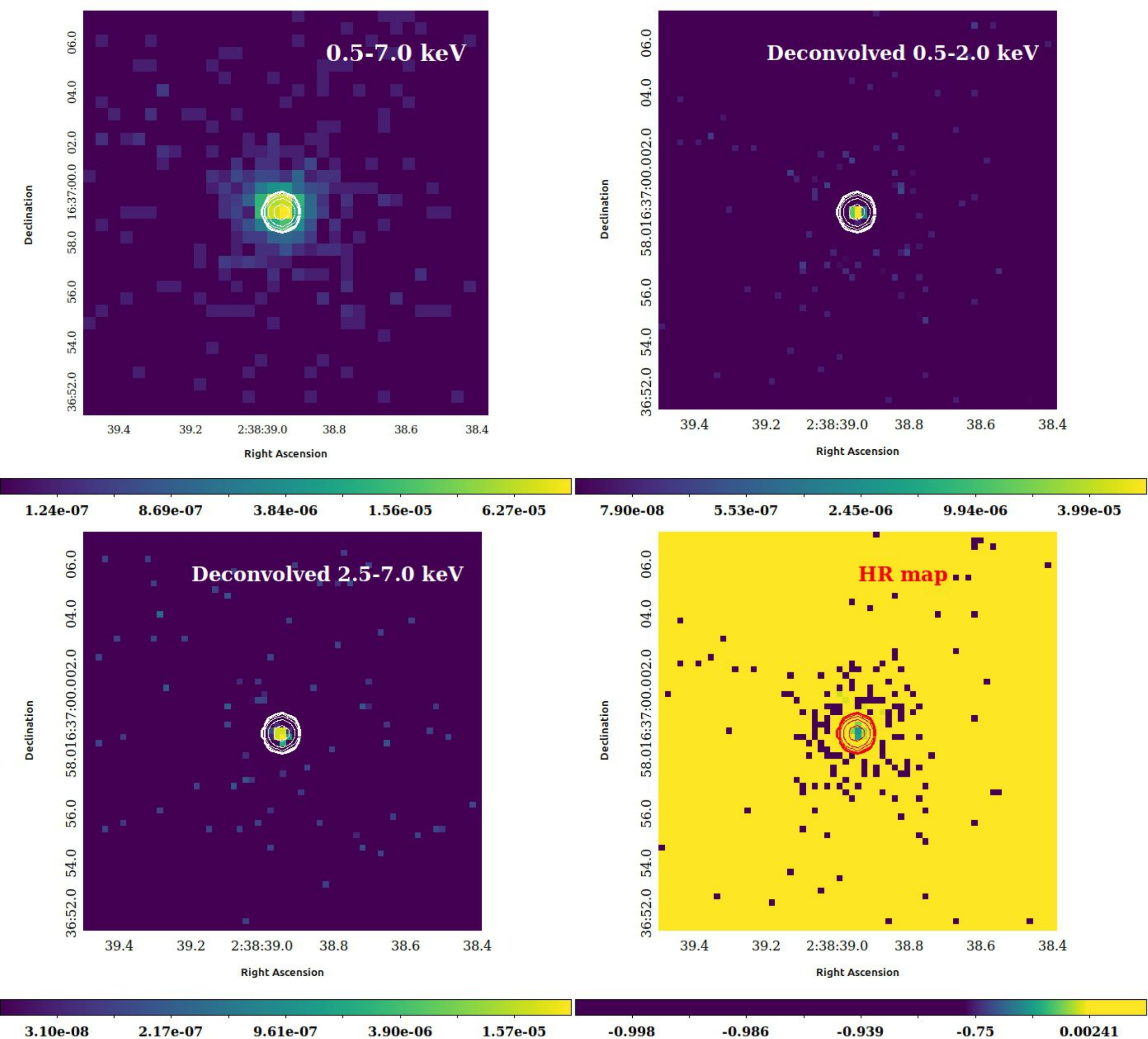}
	\caption{Low-energy (0.5--2.0\,keV) deconvolved intensity contours (at 0.5\,px resolution) of the blazar AO\,0235+16, superimposed on the event file (top left), the deconvolved soft image (top right), the deconvolved hard image (bottom left), and the averaged HR map (bottom right).} 
	\label{fig:0235+16}
\end{figure}

Based on those spectra, we performed PSF simulations, and next produced deconvolved and hardness ratio images, all as presented in Figure\,\ref{fig:0235+16}. As shown, there is no sub-structure on the hardness ratio map: a point source at the position of the blazar, characterized by HR\,$\simeq -0.5$, is surrounded by the background with the HR values of either $-1$ or 0.

The radio-loud quasar B2251+134 (=4C+13.85, $z=0.673$) has been observed (ObsID 2146) on 18-01-2000 with 25.8\,ksec exposure time. The source spectrum was extracted from a circular region (position: RA\,=\,22:54:20.9771, DEC\,=\,13:41:48.802) with a radius 7\,px ($\simeq 3.5^{\prime\prime}$), and the background annulus of 10-15\,px radii.

\begin{figure}[ht!]
	\centering
	\includegraphics[width=0.89\textwidth]{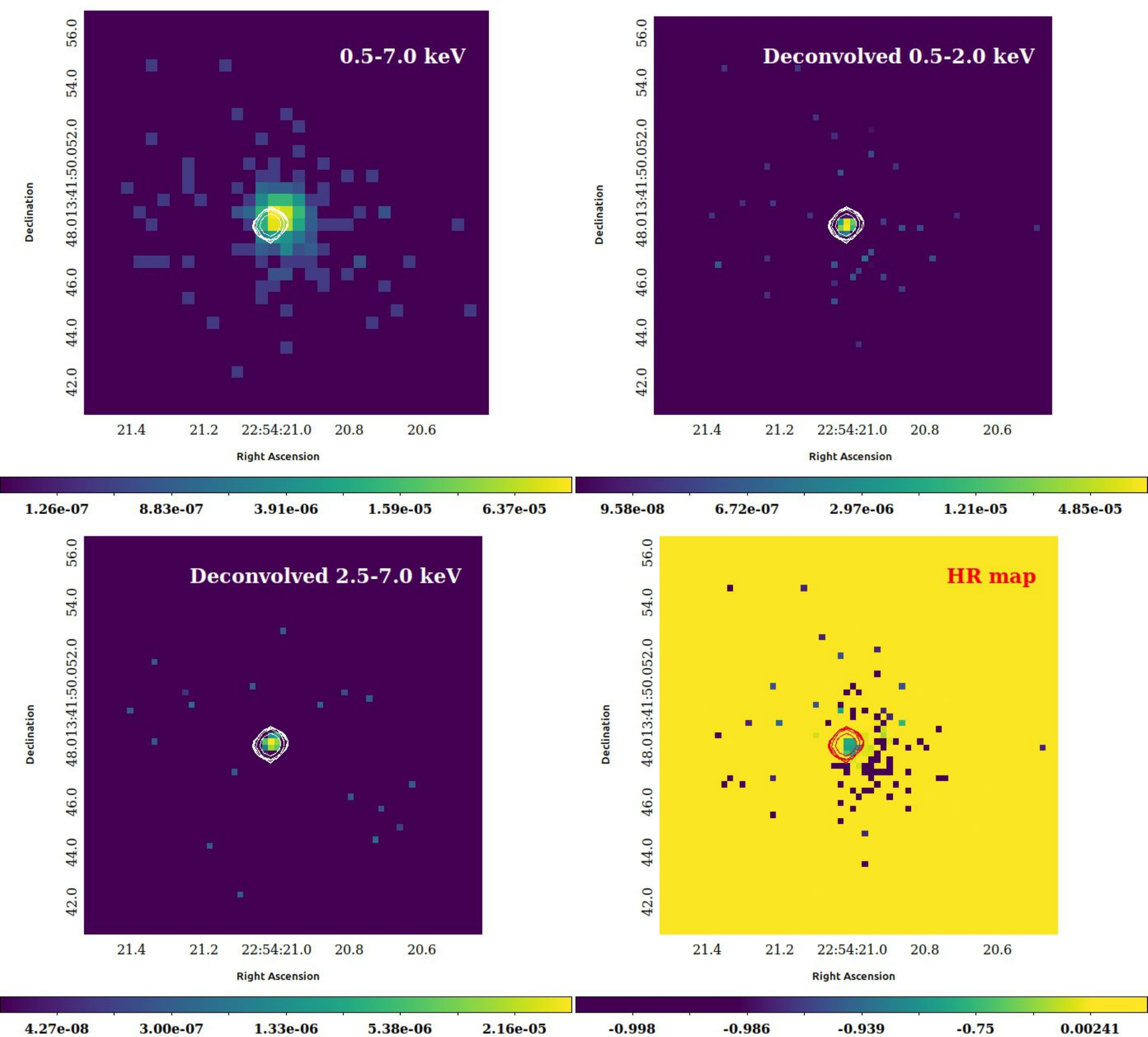}
	\caption{Low-energy (0.5--2.0\,keV) deconvolved intensity contours (at 0.5\,px resolution) of the quasar B2251+134, superimposed on the event file (top left), the deconvolved soft image (top right), the deconvolved hard image (bottom left), and the averaged HR map (bottom right).} 
	\label{fig:2251+134}
\end{figure}

We have modeled the {\it Chandra} spectra of B2251+134 in the soft ($0.5-2.0$\,keV) and hard ($2.5-7.0$\,keV) bands with single power-law models modified by the Galactic column density $N_{\rm H,\,Gal} = 4.67\times 10^{20}$\,cm$^{-2}$ \citep{HI4PI2016}, this time however allowing for the intrinsic absorption in addition to the Galactic one. The intrinsic column density was kept as a free parameter when fitting the soft spectrum; the resulting best-fit value was then frozen when fitting the hard segment of the source spectrum. The results of the following image deconvolution are presented in Figure\,\ref{fig:2251+134}. Again, what we see is a well-defined point source in the center with HR\,$\sim -0.8$, surrounded by the HR\,$=-1$ or $=0$ background.


\begin{thebibliography}{}

\bibitem[Araudo et al.(2016)]{Araudo16} Araudo, A.~T., Bell, A.~R., Crilly, A., et al.\ 2016, \mnras, 460, 3554. doi:10.1093/mnras/stw1204

\bibitem[Araudo et al.(2018)]{Araudo18} Araudo, A.~T., Bell, A.~R., Blundell, K.~M., et al.\ 2018, \mnras, 473, 3500. doi:10.1093/mnras/stx2552

\bibitem[Ba{\l}uci{\'n}ska-Church et al.(2005)]{Balucinska05} Ba{\l}uci{\'n}ska-Church, M., Ostrowski, M., Stawarz, . {\l} ., et al.\ 2005, \mnras, 357, L6. doi:10.1111/j.1745-3933.2005.08597.x


\bibitem[Blandford, \& Rees(1974)]{Blandford74} Blandford, R.~D., \& Rees, M.~J.\ 1974, \mnras, 169, 395. doi:10.1093/mnras/169.3.395

\bibitem[Brunetti et al.(2003)]{Brunetti03} Brunetti, G., Mack, K.-H., Prieto, M.~A., et al.\ 2003, \mnras, 345, L40. doi:10.1046/j.1365-8711.2003.07185.x

\bibitem[Carilli, \& Barthel(1996)]{Carilli96} Carilli, C.~L., \& Barthel, P.~D.\ 1996, \aapr, 7, 1. doi:10.1007/s001590050001

\bibitem[Carter et al.(2003)]{Carter03} Carter, C., Karovska, M., Jerius, D., et al.\ 2003, Astronomical Data Analysis Software and Systems XII, 295, 477.

\bibitem[Dabbech et al.(2018)]{Dabbech18} Dabbech, A., Onose, A., Abdulaziz, A., et al.\ 2018, \mnras, 476, 2853. doi:10.1093/mnras/sty372

\bibitem[Davis et al.(2012)]{Davis12} Davis, J.~E., Bautz, M.~W., Dewey, D., et al.\ 2012, \procspie, 84431A

\bibitem[Davis(2001)]{Davis01} Davis, J.~E.\ 2001, \apj, 562, 575.doi:10.1086/323488

\bibitem[Drake et al.(2006)]{Drake06} Drake, J.~J., Ratzlaff, P., Kashyap, V., et al.\ 2006, \procspie, 62701I. doi:10.1117/12.672226

\bibitem[Eracleous \& Halpern(2004)]{Eracleous04} Eracleous, M., \& Halpern, J.~P.\ 2004, \apjs, 150, 181. doi:10.1086/379823

\bibitem[Fan et al.(2008)]{Fan08} Fan, Z.-H., Liu, S., Wang, J.-M., et al.\ 2008, \apjl, 673, L139. doi:10.1086/528372

\bibitem[Freeman et al.(2001)]{Freeman01} Freeman, P., Doe, S., \& Siemiginowska, A.\ 2001, \procspie, 4477, 76. doi:10.1117/12.447161

\bibitem[Fruscione et al.(2006)]{Fruscione06} Fruscione, A., McDowell, J.~C., Allen, G.~E., et al.\ 2006, \procspie, 62701V. doi:10.1117/12.671760

\bibitem[Garmire et al.(2003)]{Garmire03} Garmire, G.~P., Bautz, M.~W., Ford, P.~G., et al.\ 2003, \procspie, 4851, 28. doi:10.1117/12.461599

\bibitem[Haggard et al.(2019)]{Haggard19} Haggard, D., Nynka, M., Mon, B., et al.\ 2019, \apj, 886, 96. doi:10.3847/1538-4357/ab4a7f

\bibitem[Hardcastle et al.(2004)]{Hardcastle04} Hardcastle, M.~J., Harris, D.~E., Worrall, D.~M., et al.\ 2004, \apj, 612, 729. doi:10.1086/422808

\bibitem[Hardcastle et al.(2016)]{Hardcastle16} Hardcastle, M.~J., Lenc, E., Birkinshaw, M., et al.\ 2016, \mnras, 455, 3526. doi:10.1093/mnras/stv2553

\bibitem[Harris \& Krawczynski(2006)]{Harris06} Harris, D.~E. \& Krawczynski, H.\ 2006, \araa, 44, 463. doi:10.1146/annurev.astro.44.051905.092446

\bibitem[HI4PI Collaboration et al.(2016)]{HI4PI2016} HI4PI Collaboration, Ben Bekhti, N., Flöer, L., et al.\ 2016, \aap, 594, A116. doi:10.1051/0004-6361/201629178

\bibitem[Hoshino(2008)]{Hoshino08} Hoshino, M.\ 2008, \apj, 672, 940. doi:10.1086/523665

\bibitem[Isobe et al.(2017)]{Isobe17} Isobe, N., Koyama, S., Kino, M., et al.\ 2017, \apj, 850, 193. doi:10.3847/1538-4357/aa94c9

\bibitem[Isobe et al.(2020)]{Isobe20} Isobe, N., Sunada, Y., Kino, M., et al.\ 2020, \apj, 899, 17. doi:10.3847/1538-4357/ab9d1c

\bibitem[Kataoka \& Stawarz(2005)]{Kataoka05} Kataoka, J. \& Stawarz, {\L}.\ 2005, \apj, 622, 797. doi:10.1086/428083

\bibitem[Kino, \& Takahara(2004)]{Kino04} Kino, M., \& Takahara, F.\ 2004, \mnras, 349, 336. doi:10.1111/j.1365-2966.2004.07511.x

\bibitem[Lee et al.(2011)]{Lee11} Lee, H., Kashyap, V.~L., van Dyk, D.~A., et al.\ 2011, \apj, 731, 126. doi:10.1088/0004-637x/731/2/126

\bibitem[Ligorini et al.(2021a)]{Ligorini21a} Ligorini, A., Niemiec, J., Kobzar, O., et al.\ 2021a, \mnras, 501, 4837. doi:10.1093/mnras/staa3901

\bibitem[Ligorini et al.(2021b)]{Ligorini21b} Ligorini, A., Niemiec, J., Kobzar, O., et al.\ 2021b, \mnras, 502, 5065. doi:10.1093/mnras/stab220

\bibitem[Lyubarsky(2006)]{Lyubarsky06} Lyubarsky, Y.\ 2006, \apj, 652, 1297. doi:10.1086/508606

\bibitem[Mack et al.(2009)]{Mack09} Mack, K.-H., Prieto, M.~A., Brunetti, G., et al.\ 2009, \mnras, 392, 705. doi:10.1111/j.1365-2966.2008.14081.x

\bibitem[Marchenko et al.(2017)]{Marchenko17} Marchenko, V., Harris, D.~E., Ostrowski, M., et al.\ 2017, \apj, 844, 11. doi:10.3847/1538-4357/aa755d

\bibitem[Massaro et al.(2011)]{Massaro11} Massaro, F., Harris, D.~E., \& Cheung, C.~C.\ 2011, \apjs, 197, 24. doi:10.1088/0067-0049/197/2/24

\bibitem[Massaro et al.(2015)]{Massaro15} Massaro, F., Harris, D.~E., Liuzzo, E., et al.\ 2015, \apjs, 220, 5. doi:10.1088/0067-0049/220/1/5

\bibitem[Matthews et al.(2019)]{Matthews19} Matthews, J.~H., Bell, A.~R., Blundell, K.~M., et al.\ 2019, \mnras, 482, 4303. doi:10.1093/mnras/sty2936

\bibitem[Meisenheimer et al.(1989)]{Meisenheimer89} Meisenheimer, K., Roser, H.-J., Hiltner, P.~R., et al.\ 1989, \aap, 219, 63

\bibitem[Migliori et al.(2007)]{Migliori07} Migliori, G., Grandi, P., Palumbo, G.~G.~C., et al.\ 2007, \apj, 668, 203. doi:10.1086/520870

\bibitem[Migliori et al.(2020)]{Migliori20} Migliori, G., Orienti, M., Coccato, L., et al.\ 2020, \mnras, 495, 1593. doi:10.1093/mnras/staa1214

\bibitem[Mingo et al.(2017)]{Mingo17} Mingo, B., Hardcastle, M.~J., Ineson, J., et al.\ 2017, \mnras, 470, 2762. doi:10.1093/mnras/stx1307

\bibitem[Mizuta et al.(2010)]{Mizuta10} Mizuta, A., Kino, M., \& Nagakura, H.\ 2010, \apjl, 709, L83. doi:10.1088/2041-8205/709/1/L83

\bibitem[Nandra et al.(2015)]{Nandra15} Nandra, K., Laird, E.~S., Aird, J.~A., et al.\ 2015, \apjs, 220, 10. doi:10.1088/0067-0049/220/1/10

\bibitem[Orienti et al.(2012)]{Orienti12} Orienti, M., Prieto, M.~A., Brunetti, G., et al.\ 2012, \mnras, 419, 2338. doi:10.1111/j.1365-2966.2011.19882.x

\bibitem[Orienti et al.(2017)]{Orienti17} Orienti, M., Brunetti, G., Nagai, H., et al.\ 2017, \mnras, 469, L123. doi:10.1093/mnrasl/slx067

\bibitem[Orienti et al.(2020)]{Orienti20} Orienti, M., Migliori, G., Brunetti, G., et al.\ 2020, \mnras, 494, 2244. doi:10.1093/mnras/staa777

\bibitem[O'Sullivan et al.(2018)]{OSullivan18} O'Sullivan, S.~P., Lenc, E., Anderson, C.~S., et al.\ 2018, \mnras, 475, 4263. doi:10.1093/mnras/sty171

\bibitem[Perley et al.(1997)]{Perley97} Perley, R.~A., Roser, H.-J., \& Meisenheimer, K.\ 1997, \aap, 328, 12

\bibitem[Perlman et al.(2010)]{Perlman10} Perlman, E.~S., Georganopoulos, M., May, E.~M., et al.\ 2010, \apj, 708, 1. doi:10.1088/0004-637X/708/1/1

\bibitem[Prieto et al.(2002)]{Prieto02} Prieto, M.~A., Brunetti, G., \& Mack, K.-H.\ 2002, Science, 298, 193. doi:10.1126/science.1075990

\bibitem[Pyrzas et al.(2015)]{Pyrzas15} Pyrzas, S., Steenbrugge, K.~C., \& Blundell, K.~M.\ 2015, \aap, 574, A30. doi:10.1051/0004-6361/201425061

\bibitem[Saxton et al.(2002)]{Saxton02} Saxton, C.~J., Sutherland, R.~S., Bicknell, G.~V., et al.\ 2002, \aap, 393, 765. doi:10.1051/0004-6361:20021004

\bibitem[Scheuer et al.(1974)]{Scheuer74} Scheuer P. A. G., 1974, \mnras, 166, 513. doi:10.1093/mnras/166.3.513

\bibitem[Siemiginowska et al.(2007)]{Siemiginowska07} Siemiginowska, A., Stawarz, {\L}., Cheung, C.~C., et al.\ 2007, \apj, 657, 145. doi:10.1086/510898

\bibitem[Sironi \& Spitkovsky(2011)]{Sironi11} Sironi, L. \& Spitkovsky, A.\ 2011, \apj, 726, 75. doi:10.1088/0004-637X/726/2/75

\bibitem[Stawarz et al.(2007)]{Stawarz07} Stawarz, {\L}., Cheung, C.~C., Harris, D.~E., et al.\ 2007, \apj, 662, 213. doi:10.1086/517966

\bibitem[Sunada et al.(2022a)]{Sunada22a} Sunada, Y., Isobe, N., Tashiro, M.~S., et al.\ 2022a, \mnras, 512, 5995. doi:10.1093/mnras/stac826

\bibitem[Sunada et al.(2022b)]{Sunada22b} Sunada, Y., Morimoto, A., Tashiro, M.~S., et al.\ 2022b, \pasj. doi:10.1093/pasj/psac022

\bibitem[Tavecchio et al.(2005)]{Tavecchio05} Tavecchio, F., Cerutti, R., Maraschi, L., et al.\ 2005, \apj, 630, 721. doi:10.1086/432371

\bibitem[Thomson et al.(1995)]{Thomson95} Thomson, R.~C., Crane, P., \& Mackay, C.~D.\ 1995, \apjl, 446, L93. doi:10.1086/187938

\bibitem[Thimmappa et al.(2020)]{Thimmappa20} Thimmappa, R., Stawarz, {\L}., Marchenko, V., et al.\ 2020, \apj, 903, 109. doi:10.3847/1538-4357/abb605

\bibitem[Tingay et al.(2008)]{Tingay08} Tingay, S.~J., Lenc, E., Brunetti, G., et al.\ 2008, \aj, 136, 2473. doi:10.1088/0004-6256/136/6/2473

\bibitem[Weisskopf et al.(2000)]{Weisskopf00} Weisskopf, M.~C., Tananbaum, H.~D., Van Speybroeck, L.~P., et al.\ 2000, \procspie, 4012, 2. doi:10.1117/12.391545

\bibitem[Werner et al.(2012)]{Werner12} Werner, M.~W., Murphy, D.~W., Livingston, J.~H., et al.\ 2012, \apj, 759, 86. doi:10.1088/0004-637X/759/2/86

\bibitem[Wilson et al.(2001)]{Wilson01} Wilson, A.~S., Young, A.~J., \& Shopbell, P.~L.\ 2001, \apj, 547, 740. doi:10.1086/318412

\bibitem[Xu et al.(2014)]{Xu14} Xu, J., van Dyk, D.~A., Kashyap, V.~L., et al.\ 2014, \apj, 794, 97. doi:10.1088/0004-637x/794/2/97

\end{thebibliography}
\end{document}